# Transmutation prospect of long-lived nuclear waste induced by high-charge electron beam from laser plasma accelerator

X.L. Wang,[1] Z.Y. Xu,[2] W. Luo,[1, 3, a)] H.Y. Lu,[2, b)] Z.C. Zhu,[1] and X.Q. Yan[2]

[1]*College of Nuclear Science and Technology, University of South China, Hengyang 421001, China*

[2]*State Key Laboratory of Nuclear Physics and Technology, and Key Laboratory of HEDP of the Ministry of Education, CAPT, Peking University, Beijing 100871, China*

[3]*SUPA, Department of Physics, University of Strathclyde, Glasgow G4 0NG, United Kingdom*

Photo-transmutation of long-lived nuclear waste induced by high-charge relativistic electron beam (*e*-beam) from laser plasma accelerator is demonstrated. Collimated relativistic *e*-beam with a high charge of approximately 100 nC is produced from high-intensity laser interaction with near-critical-density (NCD) plasma. Such *e*-beam impinges on a high-*Z* convertor and then radiates energetic bremsstrahlung photons with flux approaching $10^{11}$ per laser shot. Taking long-lived radionuclide $^{126}$Sn as an example, the resulting transmutation reaction yield is the order of $10^9$ per laser shot, which is two orders of magnitude higher than obtained from previous studies. It is found that at lower densities, tightly focused laser irradiating relatively longer NCD plasmas can effectively enhance the transmutation efficiency. Furthermore, the photo-transmutation is generalized by considering mixed-nuclide waste samples, which suggests that the laser-accelerated high-charge *e*-beam could be an efficient tool to transmute long-lived nuclear waste.

## I. INTRODUCTION

Thanks to current advances in laser technology, laser-plasma acceleration of electrons has been realized as a compact and bright electron source.[1-5] By focusing the laser-plasma-driven electron beam (*e*-beam) onto a high-*Z* metallic target, efficient energetic *γ*-rays can be generated through bremsstrahlung mechanism.[6, 7] As an alternative, Compton scattering of laser light with laser-accelerated *e*-beam can also be used to produce high-energy *γ*-rays.[8, 9] Radiations from both mechanisms are promising for study and applications in nuclear physics, such as radiography, medical isotope production and nuclear waste transmutation.[10-15]

The disposal of long-lived nuclear waste has become a particularly challenging issue in nuclear industry since the nuclear power is used so extensively around the world.[16, 17] Low-energy

---

a) E-mail: wenluo-ok@163.com
b) E-mail: hylu@pku.edu.cn

thermal neutrons, obtained by using medium energy intense proton beams and/or high flux nuclear reactors, are very useful for transmutations via (n, γ) reactions and/or nuclear fissions. Nuclear waste transmuted by (n, γ) reactions and those by nuclear fissions are limited to those with large neutron capture cross-sections and those with large fission branches, respectively. A lot of other fission product radionuclides, however, are produced in addition to the specific isotope of interest, such as isotopes for medical use, and thus chemical separation is required for extracting the desired isotopes. Extra separation and extraction would make nuclear waste transmutation very complicated and costly.

Besides nuclear transmutation using bombardment from medium energy proton accelerator or a fast neutron reactor, photo-transmutation induced by laser-generated electron-bremsstrahlung source should be considered as an alternative. Laser-based transmutations of $^{129}$I have already been explored experimentally,[18, 19] and those of $^{90}$Sr, $^{93}$Zr, $^{99}$Tc, $^{107}$Pd, $^{126}$Sn, $^{135}$Cs and $^{137}$Cs have been investigated theoretically.[20-27] These studies demonstrate the transmutation reaction in moderate yield of the order of $10^4 - 10^7$ s$^{-1}$ and hence suggest that laser-generated electrons (mainly due to the laser-ponderomotive acceleration) could be a promising tool for photonuclear studies as well as the resulting nuclear waste management.

In this paper, we propose an enhanced nuclear waste transmutation by using high-charge relativistic *e*-beam from laser interaction with near-critical-density (NCD) plasma. The transmutation scheme of long-lived radionuclide, $^{126}$Sn as an example, is illustrated schematically in Fig. 1. Laser-accelerated electrons are focused onto an mm-thick Ta target, as a bremsstrahlung convertor, to produce high-energy bremsstrahlung γ-rays, which in turn initiate transmutation reactions inside the Tin sample. Then the isotope $^{126}$Sn (with half-life of $10^5$ years) can be transmuted into short-lived isotope $^{125}$Sn (with half-life of 9.64 days) and stable isotope $^{124}$Sn through mainly (γ, n) and (γ, 2n) reactions, respectively. Simulations with multi-dimensional particle-in-cell (PIC) code EPOCH (version 4.8.3)[28] and general-purpose Monte Carlo code GEANT4 (version 9.3.p02)[29] are performed, respectively, to approach laser-plasma interactions and passage of particles through matter.

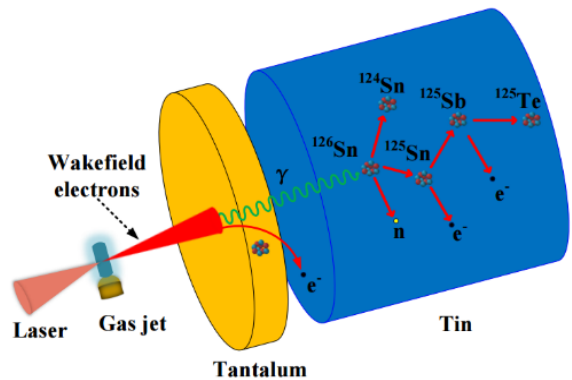

**Fig. 1.** Schematic diagram of the transmutation of radionuclide $^{126}$Sn by laser-accelerated *e*-beam. Collimated *e*-beams generated from laser interaction with gas jet are converted into high-energy bremsstrahlung γ-rays with help of tantalum target as a convertor. This is followed by the irradiation of energetic photons on tin target such that the nuclear waste can be transmuted via photonuclear reactions. The tantalum target is placed 2 cm downstream from the gas jet. Space is added between the tantalum and tin components in order to visualize clearly the target structure.

The nuclear waste transmutation driven by laser-plasma-based high-charge *e*-beam has the following merits. High-charge and collimated *e*-beam has natural advantage to produce intense medium-energy photon beam via efficient bremsstrahlung mechanism. Normally, photo-transmutation can take advantage of the resonant cross-section integrated over the whole region of GDR, which can be approximated as $0.25A$ fm$^2$ with $A$ being the mass number.[30] Furthermore, the GDR has a relatively board width (~ 5 MeV) for each individual isotope. As a result it is not needed to narrow the generated bremsstrahlung spectra. In addition, the transmutation of medium heavy nuclei is mainly induced by (γ, n) and (γ, 2n) reactions. Thus one can identify the specific / desired products and then assess their radioactivity and radio-toxicity. This is in contrast to neutron-induced transmutation, where many extra radioisotopes are necessarily produced.

Our study shows that the laser-accelerated *e*-beam with a large charge significantly enhances the transmutation efficiency. The resulting transmutation reaction yield is two orders of magnitude and even higher relative to previous achievements. This makes photo-transmutation of nuclear wastes induced by laser plasma accelerator quite promising. It should be warmly reminded that these phenomena could hardly be revealed according to the previous calculations.

## II. INTENSE BREMSSTRAHLUNG SOURCE FROM HIGH-CHARGE ELECTRON BEAM

### A. High-charge and collimated *e*-beam

Currently two main physical mechanisms, namely laser wake-field acceleration (LWFA)[31] and direct laser acceleration (DLA),[32,33] have been proposed for electron acceleration in relativistic laser-plasma interaction. The LWFA regime is promising for production of mono-energetic *e*-beam with the energies up to several GeV. However, in order to achieve long enough acceleration length before dephasing, low density plasma should be used, and the beam current is consequently limited to tens of pC. At higher plasma density, the DLA starts to come into play and even becomes dominant acceleration mechanism. Under the laser and plasma conditions given below, DLA is mainly responsible for accelerating electrons with energies lower than 30 MeV.

In our work, hybrid acceleration mechanism including both LWFA and DLA is used to produce energetic and dense *e*-beam.[34] It helps to generate a large number of bremsstrahlung photons whose spectrum fully covers the GDR energy region, and therefore results in a strong coupling of energy to the GDR for the transmutation. In contrast to DLA *e*-beams case, most electrons and photons are concentrated on low-energy region, and consequently leads to a poor coupling efficiency. Furthermore, the hybrid acceleration leads to production of a high-charge *e*-beam (with energies greater than 1 MeV) containing electrons of up to 100 nC, rather than a low-charge *e*-beam from purely LWFA. Accordingly, the potential to transmute nuclear waste should be increased drastically.

Interactions of ultra-intense laser pulses with gas jets were performed with three-dimensional (3D) PIC simulations. To optimize the electron energy and beam charge, different peak amplitude of the laser electric field $a_0$ and the plasma region length *L* are used. The laser pulse duration is fixed to 12 fs in full width at half maximum (FWHM), and the wavelength is set to 1 $\mu$m. The incident laser pulse with a Gaussian transverse field profile is focused by an f/2 off-axis parabolic mirror at the left boundary of the plasma region with a spot size of 3.6 $\mu$m (FWHM). Two cases of plasma region lengths: *L*=60 $\mu$m and *L*=90 $\mu$m are used in 3D simulations. The simulation box corresponds to a physical volume of 80 $\mu$m × 30 $\mu$m × 30 $\mu$m for the *L*=60 $\mu$m case, and 120 $\mu$m × 30 $\mu$m × 30 $\mu$m for the *L*=90 $\mu$m case. The plasma region is sampled by 12.5 cells on average per laser wavelength in laser propagation direction and 8 cells per wavelength in each transverse direction. Eight macro-electrons and two macro-protons are placed in each cell. The plasma region is placed 10 $\mu$m away from the left boundary of simulation box, with a density of $n_e = 0.5 n_c$.

Spectral distribution of laser-accelerated *e*-beams is shown in Fig. 2. Broadband *e*-beams with cut-off energies turning from 150 MeV to 250 MeV are produced under different laser and plasma conditions. At low energies, the *e*-beams have a Maxwellian-like energy distribution, which is in good agreement with that obtained with DLA mechanism[32, 33]. However, they have flat spectra at higher energies and the high-energy tails extend up to 100s MeV. It is found that the electron spectrum broadens with the laser intensity, implying more electrons are accelerated to relativistic energies. The total charge of the accelerated electrons with energies greater than 1 MeV approaches 100 nC. Considering the threshold effect for photon-induced transmutations and the threshold energy for transmutation of $^{126}$Sn is around 8 MeV, the number of electrons with energies above 6 MeV is counted. Then the consequent *e*-beam charges are about 14 and 25 nC for lasers with amplitude $a_0$ = 50 and 70, respectively, regardless of the plasma length *L* = 60 or 90 $\mu$m.

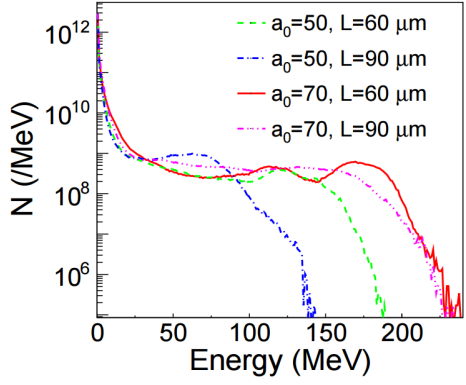

**Fig. 2.** Energy spectra of laser-accelerated *e*-beams for four different laser and plasma parameters

The *e*-beams produced from the hybrid accelerations are collimated. Fig. 3 shows transverse profiles of the *e*-beams with energies higher than 6 MeV. The collimation angles are calculated to be 100 – 300 mrad. We see that the *e*-beams produced by shorter plasmas are more collimated and have higher peak brightness. Furthermore, the collimation angle decrease with the laser intensity. This is in agreement with that shown in Fig. 2. Because the *e*-beam charge obtained with laser amplitude $a_0 = 70$ almost doubles of that with $a_0 = 50$, the *e*-beam generated with higher laser intensity is more suitable for initiating transmutation reactions.

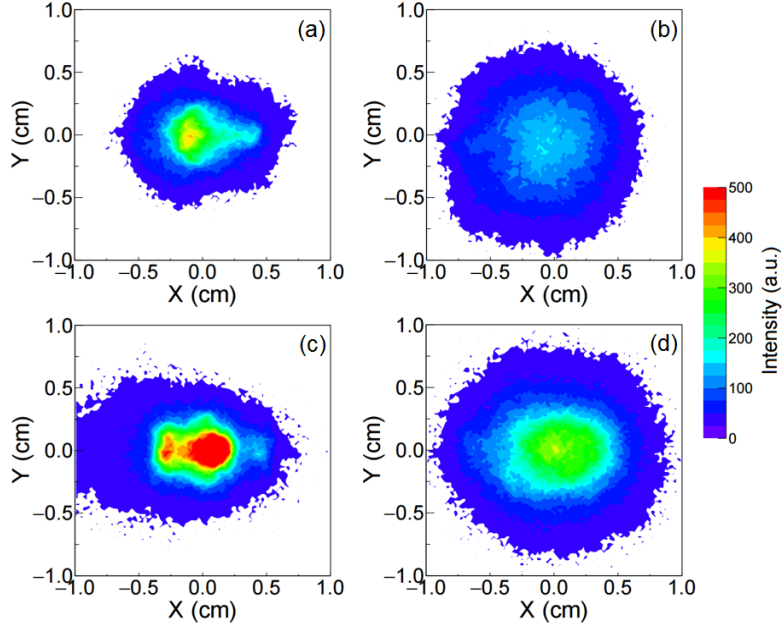

**Fig. 3.** The transverse distribution of the laser-accelerated *e*-beam recorded at 2 cm downstream of the gas jet for the laser and plasma parameters $a_0 = 50$ and $L = 60$ $\mu$m (a), $a_0 = 50$ and $L = 90$ $\mu$m (b), $a_0 = 70$ and $L = 60$ $\mu$m (c), and $a_0 = 70$ and $L = 90$ $\mu$m (d).

## B. HIGH-BRIGHTNESS BREMSSTRAHLUNG γ-RAY SOURCE

Generation of bremsstrahlung radiation and the consequent nuclear transmutations were investigated with GEANT4 simulations, in which the recommending QGSP_BERT_HP physics list is added to describe the photonuclear process.[35, 36] The QGSP_BERT_HP includes a data-driven high precision neutron package for neutrons below 20 MeV. For low and medium energy particles, it has better agreement with the experimental data in comparison with other physics lists, such as LHEP.[29] A metallic tantalum target was employed as bremsstrahlung convertor. Both the convertor and the tin sample were assumed to have cylindrical structures. Bremsstrahlung γ-rays are produced by the deceleration of energetic electrons inside the convertor. The bremsstrahlung spectra for different laser and plasma conditions are shown in Fig. 4. The photon flux increases with the laser intensity, whereas the plasma length effect is not obvious in this context. In the GDR energy region, we obtain the γ-beam flux of $(0.6 - 1.2) \times 10^{11}$ per laser shot. Such high-flux bremsstrahlung γ-ray source may attract much attention due to its diverse applications aforementioned.

The convolution of the bremsstrahlung spectrum and the reaction cross-section suggests the transmutation reaction yield. In order to reveal such convolution, also shown in Fig. 4 is the total cross-section of the dominant (γ, n) and (γ, 2n) reactions with $^{126}$Sn. Good agreement was found between the GEANT4 results and the TALYS[37] calculations for the total reaction cross-section in the GDR region, implying that GEANT4 is a reliable tool for simulations of transmutation reaction with $^{126}$Sn. The peak of the reaction cross-section ($\sigma_{peak}$) happens at the γ-ray energy of 15 MeV. The transmutation yield caused by the bremsstrahlung γ-rays shall increase with the laser intensity according to the convolution.

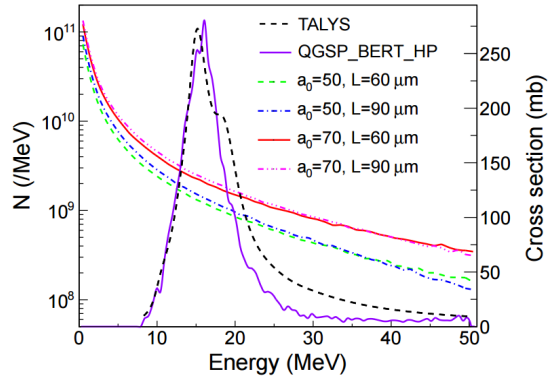

**Fig. 4.** The bremsstrahlung spectra for different laser and plasma parameters, together with the total reaction cross-section of the $^{126}$Sn (γ, n) and $^{126}$Sn(γ, 2n) reactions. The radius and the thickness of the convertor used in the simulation were $r_1$=2 cm and $T_1$=2 mm.

## III. PHOTO-TRANSMUTATION OF SINGLE-ISOTOPE SAMPLE

Long-lived nuclear waste $^{126}$Sn was chosen as an example to demonstrate the transmutation potential induced by high-energy, high-flux bremsstrahlung emissions. The significance of transmutation of $^{126}$Sn should be addressed. The $^{126}$Sn is a fission by-product of Uranium-235 (or other fissile nuclide) with the fission yield of about 0.06%. It has high radio-toxicity (6.3 Sv/g)[38] and poses a health hazard from both the beta particles and γ-rays emitted. The deposition of tin in bone and other organs and tissues will induce the risk of carcinogenesis.[24]

## A. Transmutation products

As mentioned above, $^{126}$Sn will be mainly transmuted into short-lived $^{125}$Sn through ($\gamma$, n) reaction, or stable isotope $^{124}$Sn through ($\gamma$, 2n) reaction. The $^{125}$Sn further beta decays into $^{125}$Sb, which has a half-life of 2.76 years as it decays into stable isotope $^{125}$Te by emitting an electron. Additional transmutation products are produced by other reaction channels that can be competitive with the ($\gamma$, n) and ($\gamma$, 2n) reactions. The photo-transmutation of $^{126}$Sn is summarized in Table 1. For all the reaction products, the $^{125}$Sn yield is the highest and takes a ratio of 50.3%. The following one is the $^{124}$Sn, comprising 47.2%. High-energy electrons initiate directly nuclear reaction and then contribute the transmutation with a tiny share of 0.09%. The remaining products account for only 2.5%. This is mainly caused by the large difference between their reaction cross-sections. All of the product nuclides such as $^{122}$Cd, $^{124, 125}$In and $^{123-125, 127}$Sn are short-lived or stable. Secondary neutrons and protons further induce the nuclear transmutation, thus generating a small quantity of $^{127}$Sn and $^{125}$Sb isotopes.

**TABLE 1** Photo-transmutation products of $^{126}$Sn. The laser intensity and plasma length used in the simulation were $a_0 = 70$ and $L = 60$ $\mu$m. The thicknesses of convertor target and transmuted target were 5 mm and 10 cm, respectively.

| Product nuclide | Reaction channel | Half life | Product ratio (%) | Threshold energy (MeV) | $\sigma_{peak}$ (mb) |
|---|---|---|---|---|---|
| $^{125}$Sn | ($\gamma$, n) | 9.64 day | 50.305 | 8.2 | 224.1 |
| $^{124}$Sn | ($\gamma$, 2n) | Stable | 47.208 | 14 | 135.0 |
| $^{123}$Sn | ($\gamma$, 3n) | 129.2 day | 2.147 | 24 | 13.0 |
| $^{122}$Cd | ($\gamma$, $\alpha$) | 5.24 s | 0.026 | 12 | 6.6×10$^{-5}$ |
| $^{125}$In | ($\gamma$, p) | 2.36 s | 0.031 | 14 | 0.1 |
| $^{124}$In | ($\gamma$, p + n) | 3.12 s | 0.171 | 22 | 4.6×10$^{-2}$ |
| $^{127}$Sn | (n, $\gamma$) | 2.1 hour | 0.022 | - | 4.5×10$^{3}$ |
| $^{125}$Sn | (e, n) | 9.64 day | 0.087 | - | - |
| $^{125}$Sb | (p, 2n) | 2.76 year | 0.002 | 6.7 | 1.1×10$^{3}$ |

## B. Transmutation reaction yields

Bremsstrahlung $\gamma$-rays together with the electrons that are escaped from the rear of convertor are employed to transmute the long-lived nuclear waste. Although the secondary particles produced inside the tin sample lead to additional transmutation reactions, these contributions can be neglected according to their shares listed in Table 1. Due to the collimated *e*-beam and the compact geometry, the radii of the convertor and transmuted target have insignificant effect on the transmutation yield.[26] We focus our attention on the influence of their thicknesses on transmutation yield. Fig. 5 shows the contribution of the energetic bremsstrahlung $\gamma$-rays and electrons to the transmutation reaction as a function of convertor thickness. The contribution of the bremsstrahlung $\gamma$-rays first increases and then decreases with the convertor thickness. The electrons result in decreased transmutation yields. Such decrease is mainly caused by particles absorption inside the convertor. It is found that the total transmutation yields have peak values as the convertor thicknesses are 3 mm and 5 mm for laser amplitude $a_0$ = 50 and 70, respectively.

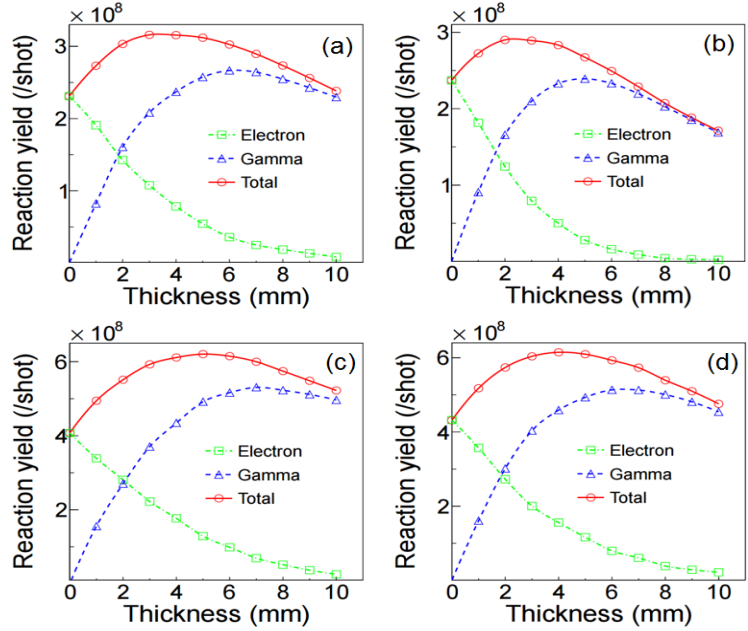

**Fig. 5.** The contribution of the electrons and bremsstrahlung $\gamma$-rays to the transmutation reaction for laser and plasma parameters $a_0$ = 50, $L$ = 60 $\mu$m (a), $a_0$ = 50, $L$ = 90 $\mu$m (b), $a_0$ = 70, $L$ = 60 $\mu$m (c) and $a_0$ = 70, $L$ = 90 $\mu$m (d). The total contributions are also shown in the figures. Both the radius and thickness of the tin sample used in the simulations were 2.0 cm.

By using the optimized convertor thickness, the dependence of transmuted target thickness on the reaction yield was investigated and is shown in Fig. 6. The reaction yield increases with the thickness of tin target. For a given laser amplitude, the plasma length can hardly affect the transmutation reactions. Considering the transmutation efficiency per unit volume of the transmuted target, the thickness of the transmuted target is suggested to be 5 cm. Accordingly the yields of the transmutation reactions reach 5 × $10^8$ and 1.2 × $10^9$ per shot for lasers with peak amplitude $a_0$ = 50 and 70, respectively. The $e$-beam charge mainly results in the difference of the reaction yields.

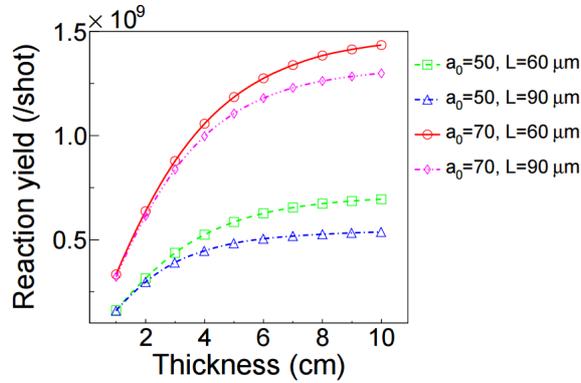

**Fig. 6.** The reaction yield as a function of the thickness of transmuted target under four different laser and plasma conditions. The optimized thicknesses of the convertor shown in Fig. 5 were used for their respective simulations.

## C. Effect of laser and plasma conditions

We study the effect of laser and plasma conditions on the transmutation reactions. 2D PIC simulations have been performed with a wider range of laser spot sizes varying from 4 to 20 $\mu$m. Meanwhile, the laser peak power is kept at 1 PW with a duration of 30 fs, which is achievable by state-of-art laser technology, and the wavelength of laser is set to 1 $\mu$m. Plasma lengths $L$ = 30, 60 and 90 $\mu$m and plasma densities $n_e$ = 0.3$n_c$, 0.5$n_c$ and 0.7$n_c$ are used in these simulations. The 2D simulation box is sampled by 6000 cells in laser propagation direction and 3200 cells in transverse direction, which corresponds to a physical volume of 150 $\mu$m × 100 $\mu$m. The plasma region is placed 10 $\mu$m away from the left boundary of simulation box, and the laser is focused at the left boundary of the plasma region. Sixteen macro-electrons and two macro-protons are placed in each cell.

Fig. 7 shows the dependence of the reaction yield on these laser and plasma parameters. It is found that the transmutation reaction yield decreases rapidly with the spot size. This is mainly caused by the decrease of the laser-accelerated $e$-beam charge as well as its cut-off energy. At lower densities, the relatively longer plasmas, i.e. $L$ = 60 or 90 $\mu$m, can result in production of more energetic electrons, thus enhance the transmutation efficiency. This is also correlated to the $e$-beam charge and cut-off energy. In addition, as the laser spot size attains a certain value, both the plasma density and length can hardly affect the reaction yield. On the whole, our study shows that at lower densities, tightly focused lasers interactions with relatively longer plasmas are beneficial to nuclear waste transmutation.

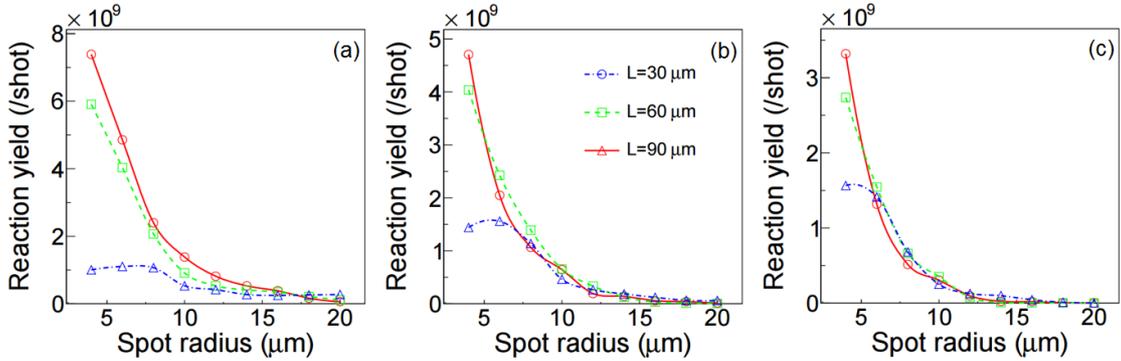

**Fig. 7.** The reaction yields as a function of laser focus for plasma density of 0.3$n_c$ (a), 0.5$n_c$ (b) and 0.7$n_c$ (c) considering that the laser peak power is 1 PW. In the simulations, the radius and thickness of the convertor were 2.0 cm and 5 mm, and those of the tin sample were 2.0 cm and 5 cm, respectively.

## IV. PHOTO-TRANSMUTATION OF MIXED-NUCLIDE WASTE SAMPLES

To generalize the nuclear waste transmutation induced by the high-charge relativistic $e$-beam, we quantify the implication of mixed-nuclide waste samples consisting of four main long-lived radionuclides $^{93}$Zr, $^{107}$Pd, $^{126}$Sn and $^{135}$Cs for the final reaction rate, rather than assuming the single-isotope sample $^{126}$Sn. These radionuclides with intermediate mass have considerable reaction cross-sections in the GDR region. Fig. 8 shows the atomic mass $A$ versus atomic number $Z$ of their transmutation fragments. One can see that the ($\gamma$, n) and ($\gamma$, 2n) reactions play dominant role in nuclear waste transmutation and the reaction types of $^{93}$Zr, $^{107}$Pd and $^{135}$Cs are similar to that of $^{126}$Sn. The transmutation yield for each of these radionuclides is a few times $10^8$ per laser shot and the total one exceeds $10^9$ per laser shot for laser with peak amplitude of $a_0 = 70$. The transmutation capability for the mixed-nuclide waste samples case can therefore be comparable with that for the single-isotope sample case.

The irradiation time has direct effect on the transmutation yield. Considering the laser repetition rate of 1 Hz, it would be possible to induce about $4 \times 10^{12}$ reactions after one hour irradiation. Previous studies have demonstrated that for the single-isotope sample $^{126}$Sn, only about $2.7 \times 10^{10}$ reactions per hour were achieved with a laser system of $10^{21}$ W/cm$^2$ and the repetition rate of 100 Hz.[24] Regarding transmutation of $^{135}$Cs, $^{107}$Pd and $^{129}$I based on laser-Compton scattering using 1 kW laser system, the yield of transmutation reactions was $10^8$ – $10^{11}$ per hour.[8, 12, 23] The comparison shows that the transmutation yield obtained in this system is two orders of magnitude higher than those achieved in previous studies. This suggests great prospective potential in transmutation of long-lived fission radionuclides using high-charge relativistic $e$-beams from laser plasma accelerator.

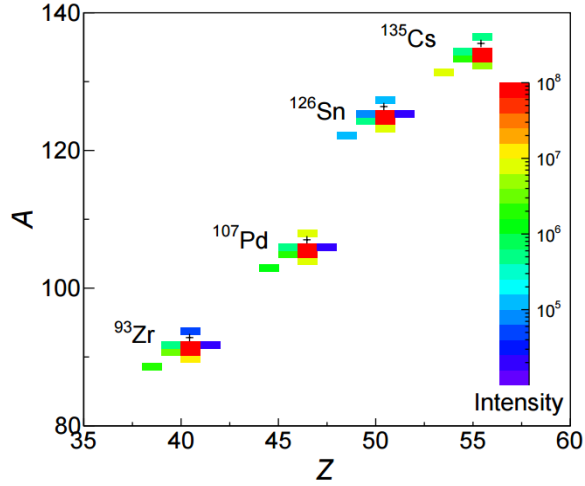

**Fig. 8.** Atomic mass $A$ versus effective nuclear charge $Z$ of the transmutation fragments of long-lived radionuclides $^{93}$Zr, $^{107}$Pd, $^{126}$Sn and $^{135}$Cs. The components of these mixed-nuclide samples have the same weight. The parameters used for calculation are laser intensity of $a_0 = 70$ and plasma length of $L = 60$ $\mu$m, and the target sizes are the same as Fig. 7.

## V. SUMMARY

Considering the single-isotope and mixed-nuclide waste samples cases, nuclear waste transmutations driven by laser-accelerated relativistic $e$-beam have been investigated through PIC and Monte-Carlo simulations. It is shown that collimated relativistic $e$-beam with a high charge of approximately 100 nC, produced by high-intensity laser irradiating near-critical-density plasma, can generate high-flux ($10^{11}$ per laser shot) bremsstrahlung γ-rays in the GDR energy region. The transmutation reaction yield is obtained to be the order of $10^9$ per laser shot, which is two orders of magnitude higher than obtained in previous studies. Moreover, it is found that tightly focused lasers impinging on longer NCD plasmas at lower densities can enhance the transmutation yield. In view of the current advances in tabletop ultra-intense lasers, the proposed scheme for enhanced transmutation reactions is promising for nuclear waste management and medical isotope production.

## ACKNOWLEDGMENTS

This work was supported by the National Natural Science Foundation of China (Grant Nos. 11405083, 11675075, 11575011 and 11605084) and the Graduate Student Innovation Project Foundation of the University of South China (Grant No. 2016XCX22). W.L. appreciates the support from the Young Talent Project of the University of South China.